# VARIABLE-PHASE ASYNCHRONOUS CYCLOTRON

A. R. Tumanyan, E. Zh. Sargsyan and Z. G. Guiragossian*

*Yerevan Physics Institute (YerPhI), Yerevan, ARMENIA 375036*
*\*Guest Scientist at YerPhI*

## ABSTRACT

The conceptual design of a Variable-Phase Asynchronous Cyclotron (VPAC) is describe, which provides longitudinal bunch compression of accelerated proton or ion beams, and thus, permits high current acceleration at higher accelerator efficiency, where the possible beam losses are minimized and the accelerator's mechanical tolerances are relaxed. Beam control is assured by the ability to independently set and vary the acceleration phase and rf voltage amplitude, the inter-cavity harmonic number and the transverse focusing strength, which considerably overcome the space charge effects in each sector and turn of the proposed cyclotron. The new accelerator concept is especially suitable to accelerate intense proton beams up to 800 MeV in energy and average beam current in the 100-mA class. All accelerator elements are based on currently available and feasible technologies. To demonstrate feasibility of design, the detailed calculations and modeling of a 10-turn VPAC prototype for the production of 25.6 MeV, 100 mA proton beam are presented and the key features of the new accelerator concept are discussed.

Corresponding authors: A.R.Tumanyan, Yerevan Physics Institute, Alikhanian Broth. St.2,
Yerevan 375036, Armenia
Phone: (374 1) 355232
E-mail: tumanyan@uniphi.yerphi.am
tumanyan@lx2.yerphi.am
E.Zh.Sargsyan, Yerevan Physics Institute, Alikhanian Broth. St.2,
Yerevan 375036, Armenia
Phone: (374 1) 737411
E-mail: edgar@jerewan1.yerphi.am

# INTRODUCTION

The Variable-Phase Asynchronous Cyclotron (VPAC) is similar to the operation of a rf linear accelerator wrapped into a spiral, in which the equilibrium phase of acceleration $\varphi_e$, the acceleration voltage amplitude in consecutive cavities $U_n$ and the beam path length between cavities $S_n$, can vary individually, independent from each other. These conditions physically cannot be created in classical ring Isochronous Cyclotrons (IC) [1] or conventional Separate Orbit Cyclotrons (SOC) [2,3,4]. This is because in such cyclotrons, common sector cavities span the radial extent of turns, with multi-turn channel gaps, in which $\varphi_e$, $U_n$ and $S_n$ must be kept fixed to satisfy the isochronism condition. However, variable mode operations in $\varphi_e$, $U_n$ and $S_n$ can be created in a modification of the separate orbit Asynchronous Cyclotron (AC) concept [5,6], namely, in the VPAC approach, which is now presented and elaborated. Here, the term asynchronous is used to mean the non-isochronous operation in cyclotron geometry.

The concept of AC is best described as a modification of the operating parameters of a SOC, having the same external structure, as seen in Figure 1. If both, the turn radius $R_n$ and the acceleration voltage frequency $f$ in a SOC are selected to be large, such that $q$, the inter-cavity harmonic number, is a large integer at injection:

$$q = \frac{h}{N_c} = \frac{2\pi R_n f}{\beta_n c N_c} \tag{1}$$

where $N_c$ is the number of cavities in a cyclotron stage and $\beta_n c$ is the speed of particles, then $q$ would decrease continually in the course of acceleration, as the speed of particles increased from turn to turn.

Instead of the synchronous condition for all orbits in an isochronous cyclotron, which requires the harmonic number $h = f_{rf}/f_{rev} = 2\pi R_n f/(\beta_n c)$ to be a constant integer (constant rotation period for all turns), in an AC, the inter-cavity harmonic number $q = h/N_c$ is also required to be an integer, but may change in discrete steps in the acceleration process. The reduced harmonic number $q$ needs to remain at a specific constant value only between consecutive acceleration gaps. In an asynchronous cyclotron, hopping over discrete integer values of $q$ significantly changes the bunch revolution frequency in each turn, which is accomplished by appropriately designing and changing the magnetic path length in each sector. Thus, by necessity, the asynchronous cyclotron concept can be applied only in separate-orbit cyclotrons.

As initially discussed [5,6], the asynchronous cyclotron concept included the following first four features.

First, this scheme permits to limit the growth of orbit separations between the injection radius $R_{inj}$ and the radius at extraction $R_{ext}$, while allowing the average radius $R_{ave}$ of the cyclotron to be sufficiently large.

Second, the turn-to-turn separation for all orbits can now be designed to have a nearly equal and suitably large value, to provide for the use of a large radial aperture vacuum chamber, for the acceleration of high current beams at small losses.

Third, in a large $R_{ave}$ AC, sector magnets have much reduced field strengths, removing the necessity to use superconducting magnets.

Fourth, the length of straight sections between sector magnets is increased, to contain much stronger transverse focusing elements.

In this paper additional innovations are made to increase the utility and advantage of asynchronous cyclotrons, for which the new accelerator concept is named the Variable-Phase Asynchronous Cyclotron—VPAC, according to the following next five features.

Fifth, the bunch acceleration equilibrium phase $\varphi_e$ can be varied over a wide range, individually in each sector, by also modifying the beam path length between adjacent cavities $S_n$.

Sixth, the acceleration voltage amplitude $U_n$ in each sector can be varied independently, by mechanical [2] and electrical means introduced in cavity channels, which will be described in a separate publication.

Seventh, consequently, longitudinal bunch compression can now be programmed in each sector, by suitably selecting $\varphi_e$ and $U_n$ values. This feature increases the acceleration efficiency, reduces beam losses due to increasing longitudinal bunch size as it occurs in isochronous cyclotrons.

Eighth, the ability to independently vary the transverse focusing strength in all sectors and turns compensates the space charge dependent, betatron oscillation tune shift, $\Delta Q$ and avoids the resonances.

Ninth, the ability to independently vary and set the basic parameters of $\varphi_e$, $U_n$, $\Delta Q$, relaxes the required mechanical tolerances in all sectors of the cyclotron.

The principle of operation of such an accelerator consists of the following (see also Figure 1).

We note that in injected beams and those from strongly focused magnetic lines, usually particles at the head of a bunch come with higher energy than particles at the center; and those at the tail of the bunch have the lower energy.

As this type of injected beam enters the cyclotron, at the first acceleration gap of the first cavity, the equilibrium phase $\varphi_{e1}$ for the acceleration of bunches is selected with the help of the rf generator setting. Also, the acceleration voltage amplitude $U_1$ is set, based on the beam parameters, the follow-on beam path length $S_1$, the momentum compaction factor and the other features in the first sector of the cyclotron's first turn.

The selection of $\varphi_{e1}$ and $U_1$ is made such that upon exiting from the first cavity, the particle distribution in bunches would be reversed—particles with higher energy would now appear in the tail of bunches instead of being initially at the head. The strength of this reversal must be such that the fast particles in the tail, just prior to entering the second cavity after a path length $S_1$, would catch up with or overtake the slow particles in the head of the bunches. This produces the desired longitudinal bunch compression at the second sector. Similar procedures are used in successive sectors and turns.

We also note that all sectors are essentially non-isochronous (asynchronous) transport lines, in which phase-space rotation of the longitudinal emittance ellipse is occurring [7]. The value of the equilibrium phase in subsequent acceleration gaps, on the average, will be increased gradually, approaching the wave crest at 0 degree, except for the small phase decreases to compensate for the expansion of the bunches. The above makes it possible to effectively increase the acceleration efficiency and accordingly to reduce the number of beam turns.

Bunch processes take place in six-dimensional phase space, under the validity of Liouville's theorem, and beam parametric changes can be estimated as solutions of Vlasov's equation [7]. As a result of the forced variation of the bunch length, it is now possible to have a reduction of the longitudinal emittance, coupled with increasing the transverse emittances, while the six-dimensional phase space density is preserved. The resulting increase in the horizontal emittance will be more than the increase in the vertical emittance, as will be shown in the detailed numerical calculations and modeling of the 25.6-MeV prototype accelerator.

The above considerations indicate that for such an accelerator it will also be necessary to have large turn-to-turn separations, typically in the range of 24 – 34 cm, to accommodate for the acceleration of 100-mA class beams. The vacuum chamber of the separated orbits would have a vertical aperture of 5 – 7 cm and a horizontal aperture of 10 – 14 cm, provided that a small-emittance modern injector, such as the radio frequency quadrupole (RFQ) linac [8] is used. Modern computer-code calculations with ability to handle the bunch space charge in three dimensions, such as the Los Alamos National Laboratory's Trace-3D Beam Dynamics Program [9] is employed in this paper, to demonstrate feasibility of this accelerator's conceptual design approaches.

At lower energies of protons or ions rapid changes in $\beta$ take place, to be able to manipulate the hopping of inter-cavity harmonic number $q$ over integer values. Since it is also desirable to maintain the accelerator radius of a cyclotron stage at a reasonable value, it appears from Equation (1) that the VPAC accelerator scheme can be applied for proton energies of up to 800 MeV, for the acceleration of 100-mA class beams, in a few successive stages. All accelerator elements are based on currently available and feasible technologies. Similarly, this concept can act as a powerful injector, in the size given in the prototype's numerical example, to inject into other types of higher energy proton or ion accelerators.

## CONDITIONS FOR BUNCH COMPRESSION IN THE VPAC

The creation of bunch compression in this accelerator scheme is inherently made possible by (a) the available independent setting of the rf acceleration equilibrium phase $\varphi_e$, in each acceleration gap, and (b) the available independent setting of the acceleration voltage amplitude $U_n$, in any n-th sector acceleration channel. In this case, the continuous differential equation for synchrotron oscillations does not apply. Consequently, at start up, $\varphi_e$ in each cavity gap is determined by the following considerations.

Upon exiting an rf acceleration gap, the energy gain of an extreme particle $a$ at the head of a bunch $\Delta E_a$, the energy gain of an extreme particle $b$ at the tail of a bunch $\Delta E_b$, and the energy gain of the equilibrium particle in a bunch $\Delta E_e$ are obtained by

$$\Delta E_a = U_n T_z \cos(\varphi_e - \Psi) \quad \text{and} \quad \Delta E_b = U_n T_z \cos(\varphi_e + \Psi) \tag{2}$$

$$\Delta E_e = U_n T_z \cos \varphi_e \tag{3}$$

where $T_z$ is the transit time factor and $2\Psi$ is the full phase width of the bunch.

The transit-time factor $T_z$ is determined by

$$T_z = \frac{\sin \Delta \Phi / 2}{\Delta \Phi / 2} \tag{4}$$

where

$$\Delta \Phi = \frac{2\pi f_{rf} L_{gap}}{\beta c} \tag{5}$$

and $L_{gap}$ is the full acceleration gap in the rf cavities. Normally, $T_z$ can be maintained at a constant value if the acceleration gap is increased in proportion to the speed of the accelerated beam, which in turn increases the amplitude of the acceleration voltage $U_n$, for the same electric field in the sector cavities, to achieve a specified energy gain.

If $E_e$ is the energy of the synchronous particle as the bunch enters a cavity gap and $\Delta E_s$ is the half width of the energy spread in the bunch, then upon exiting the cavity, the energy of the extreme particles at the head ($E_a$) and the tail ($E_b$) will be $E_a = E_e - \Delta E_s + \Delta E_a$ and $E_b = E_e + \Delta E_s + \Delta E_b$. Consequently, $\varphi_e$ can be determined in each resonator and turn from

$$\sin \varphi_e = \frac{E_a - E_b + 2\Delta E_s}{2 U_n T_z \sin \Psi} \tag{6}$$

Setting $E_a = E_b$ provides the condition for monoenergetic bunches. However, the equilibrium phase for acceleration is also set within the limits of

$$(\Psi - \pi/2) < \varphi_e < 0 \tag{7}$$

such that the phase of off-center particles always remains on the rising side and below the crest of the rf wave, to maintain a Gaussian bunch distribution.

The bunch duration $\tau_f$ at the end of any sector (at the space located between two adjacent cavities) is given by

$$\tau_f = \tau_s + \Delta \tau \tag{8}$$

where

$$\Delta \tau = \frac{S_n}{c}\left(\frac{1}{\beta_b} - \frac{1}{\beta_a}\right) \tag{9}$$

$\tau_s = 2\Psi T/(2\pi)$ is the initial bunch duration, $S_n$ is the orbital path length of particles in the n-th sector and T is the rf period duration.

In Equation (9), if a positive sign is obtained for $\Delta \tau$, which is when $\beta_a > \beta_b$, bunch elongation and an increase in $\tau_f$ is described; if a negative sign is obtained for $\Delta \tau$, which is when $\beta_a < \beta_b$, axial bunch compression with a corresponding decrease in the value of $\tau_f$ is described.

The last case is possible only when a head-to-tail particle reversal configuration is obtained. In addition to the independent setting of $\varphi_e$ and $U_n$ in each sector, use of the particle head-to-tail reversal configuration is the third condition to support the axial compression of bunches in this accelerator.

In Equation (8), a negative sign for $\tau_f$ signifies the over-compression of bunches, after which, a normal distribution of particles is restored with the higher energy particles appearing at the head.

If conditions are found by the proper selection of $\varphi_e$ to always support the reverse-particle distribution, $\tau_f$ will continually decrease. From Equation (9) it is evident, to obtain large and negative values of $\Delta \tau$, it is also necessary to have large sector path lengths $S_n$ and large negative differences in the reciprocals of $\beta$. These are the fourth and fifth conditions to achieve axial bunch compression in this accelerator.

The values of beam injection energy $E_{inj}$, the transverse and longitudinal emittances at injection, the initial bunch duration $\tau_{s,i}$, the injection radius $R_{inj}$, the number of acceleration cavities $N_c$, the maximum amplitude of acceleration voltage $U_n$, and the rf frequency $f$, are normally chosen on the basis of available and feasible technologies. Having selected these parameters beforehand, and based on the simultaneous solution of Equations (3) – (9), the required value of the acceleration equilibrium phase $\varphi_e$ is determined, such that bunch compression in a given sector is achieved. The analytical solution of these coupled equations is sufficiently cumbersome to reproduce here. A computer-code calculation provides the optimized values of the above parameters, which are then inserted into the Trace-3D [9] computations for the modeling of each sector.

In this accelerator concept, the following is the control algorithm for the steering of bunches from one sector to the next. As a function of the injected beam's measured energy spread $\Delta E_{s,i}$, the duration of bunches $\tau_{f,i}$, the design value of the beam path length in the first sector $S_1$, and the selected rf acceleration voltage in the first resonator $U_1$, the rf phase is set at the generator. This is done such that an equilibrium phase $\varphi_e$ is produced on the rising side of the acceleration voltage, which after the passage of a bunch in the first rf gap, will cause at once the inversion of particles.

That inversion should be of sufficient strength to reduce the duration of bunches $\tau_{f1}$ to a small value, at the end of the beam path in the first sector. Thus, there are two cases to consider; one, which conserves the inverted distribution and the other, which induces inversion and over-compression and then restores back to the normal distribution of particles in a bunch.

In the first case, when it is necessary to obtain a monoenergetic beam, the equilibrium phase in the last resonator channel is set, such that the acceleration takes place on the falling side of the rf field. In this case $\Delta E_s$ in Equation (6) takes on a negative sign, setting $E_a = E_b$. This condition is desirable when the beam is just being extracted from the accelerator. In the second case, instead, the rising side of the rf voltage is used to obtain bunches of small duration. For the most desired case of preserving the inverted distribution of particles, it is necessary to work only on the rising side of the rf acceleration field.

The path length of particles in sectors is set by the parameters of the bending magnetic system, which essentially must change from sector to sector, to provide the desired values of both $q$ and $\varphi_e$. To have the possibility of precise tuning and to relax the maintenance of mechanical tolerances of the accelerator components and their alignment, different correction elements will be placed in the straight section of each sector. This is in addition to the quadrupole lenses for the strong transverse focusing of the beam and a number of beam-monitoring elements. In particular, wiggler type chicane magnets will be installed in the straight sections to adjust the path length of particles at the required values. Finally, it is estimated that by having large turn-to-turn separation and the other features of the VPAC, the mechanical tolerances of accelerator elements and their alignment are relatively relaxed by an order of magnitude and need be only in the order of $10^{-3}$.

The important relaxation of tolerances in this accelerator concept is one of its main advantages, in comparison to other similar accelerator structures, such as the isochronous separate orbit cyclotrons [2 – 4]. In the latter, the necessity of strictly maintaining the isochronism of particle motion reduces to having tight tolerances, which in practice are difficult to implement. Other important advantages are due to the features of longitudinal bunch compression and strong transverse focusing. These make it possible to accelerate bunches at an equilibrium phase close to the wave crest at 0 degree, which in turn, increases the efficiency of acceleration, decreases the number of turns, decreases the beam losses, and increases the number of accelerated particles in bunches.

Basically, this accelerator concept's deficiency is the uniqueness or unprecedented nature of the sector bending magnetic system. This complicates the standardization of their manufacture and tuning. However, some technical innovations already made, facilitate the solution of these problems. This concerns the fabrication of magnet yokes from iron sheets with the ability of mechanically changing the magnetic lengths and the remote control of the magnetic alignment in each sector and turn. Also, the variety of magnetic path lengths required in the sectors of each turn can be accommodated by the use of several short standard dipole modules of a few types, which are appropriately arranged. The individual supply and control of the sector bending magnets and the quadrupole focusing lenses using modern electronics and computers is straightforward to implement.

The strong beam transverse focusing elements in the straight section of each sector are of the type normally found in strong focusing synchrotrons. In particular, the separate function periodic magnetic structure can be of the FODO type. In this accelerator concept the main difference will be the possibility of having a slowly varying betatron oscillation frequency, in going from one focusing period to another. This will compensate the frequency shift of the betatron oscillations, due to space charge and other effects.

## RESULTS OF NUMERICAL CALCULATIONS AND MODELING

Numerical calculations and modeling based on the Trace-3D [9] and other computer codes are performed for a prototype VPAC accelerator. This is done to show the feasibility of conceptual design approaches and to identify the accelerator's main components, employing the available and feasible technologies. A 100 mA average current proton beam from a 2.0-MeV RFQ linac is injected into the designed accelerator structure, in which four rf acceleration cavities operate at 50 MHz. The operating frequency of the RFQ injector could be higher (i.e. 350 MHz), for which a bunch manipulation rf scheme is used to convert the frequency of the injected bunch train to 50 MHz, which will be described in a separate publication. For the injected beam, a longitudinal emittance of $1.0\pi$ degree-MeV is assumed and a high transverse emittance of $25.0\pi$ mm-mrad is taken.

The drawing in Figure 1 shows the first three turns beam orbits of the asynchronous prototype cyclotron, in which a proton beam is accelerated from 2 MeV to 25.6 MeV in 10 turns. The accelerator has $N_c$ = 4 cavities, each with 10 beam channels, and 8 sectors per turn (a total of n = 80 sectors and 40 independently tuned inter-cavity sections). Beams are injected at a radius of $R_{inj}$ = 2.5 m and extracted at $R_{ext}$ = 5.5 m, where the turn-to-turn separation is in the range of 0.24 – 0.34 m.

The prototype accelerator's conceptual design is also based on the use of modern room-temperature rf acceleration cavities similar to those developed at the Paul Scherrer Institute [10] and elsewhere [10], for operation in the range of 40 – 50 MHz. These would have a length of 6.0-m, height of 3.0 m and width of 0.3 m and sustain a peak voltage of 1.1 MV. The radial extent of the useful beam channel for the 10 turns would be 4.0 m long.

Key parameters of the Variable-Phase AC prototype accelerator are given in Table 1. Similarly, the geometrical layout of the acceleration gaps, the sector bending magnets and quadrupole focusing lenses and drift sections are shown in Figures 3 – 4, only for the first and tenth turns. However, all 10 turns were calculated and modeled. In these Figures, Trace-3D numerical calculation results are shown for the key beam dynamical parameters, including changes in the transverse-longitudinal emittances, the beam envelopes, the bunch length and the actual size of the beam in three dimensions.

Comparing the beam envelopes in Figures 3 – 4, it is seen that the accelerated beam is well confined, within the design's tolerable limits. A beam vacuum vessel of 7 cm in height and 14 cm in width for the separated turns can easily be accommodated inside these magnets, and within the turn-to-turn separation of 24 – 34 cm. This ensures that beam losses will be minimal and within the requirement of less than 0.1 nA/m.

The additional advantage of having axial bunch compression in VPAC accelerators is best seen by comparing the energy spread of the accelerated beam at the end of the first turn and at extraction, in Figures 3 and 4. While at the end of the first turn, the energy spread of the beam is 2.75% HWHM, this is reduced to $\Delta E_s / E_{ext} = 0.30\%$ HWHM at the end of the tenth turn, when the beam will be extracted.

The variation of important parameters in the course of acceleration from 2.0 to 25.6 MeV is shown in Figures 2 and 5 – 10. Figure 2 displays the designed variation of the inter-cavity harmonic number $q$, over the 40 sections in the course of acceleration in the prototype VPAC. Figure 5 shows the corresponding changes in the equilibrium phase of bunches $j_e$. Figure 6 gives the settings of the acceleration voltage amplitude $U_n$. Figure 7 is the settings of the inter-sector path length $S_n$. Figure 8 gives the settings of the sector magnetic path length $L_m$. Figure 9 displays the corresponding settings of the sector magnetic dipole field strength $H_m$. Figure 10 gives the energy of the equilibrium particle in a bunch $E_e$.

In this accelerator concept, sector magnets have low field strengths, in the range of 0.39 – 0.72 T. As the length of the sector bending magnets varies from 0.34 m to 1.0 m over the ten turns, smaller dipole magnetic modules of a few types will be used to assemble sector magnets of different lengths, for each turn. The use of small modular optimized magnets will also permit to standardize the fabrication of all sector magnets, reducing cost and fabrication time. The remaining free straight-section lengths, after the placement of acceleration cavities and the bending magnets, will be in excess of 2.0 m per sector. This is more than sufficient for the placement of eight strong focusing quadrupole lenses and beam diagnostic detectors, and to ensure the 100% extraction of the beam in the last turn.

## CONCLUSION

We have shown that the VPAC accelerator concept combines the most desirable features of the highest-current-producing proton rf linear accelerators and the compact, efficient cyclotrons. The feasibility of the conceptual design approach has been demonstrated by the use of three-dimensional space-charge numerical calculations and modeling. The presented prototype 25.6-MeV, 100-mA proton VPAC accelerator has several applications, in stand-alone mode, or as a high-current injector to other types of proton or ion accelerators, or in extended mode to produce intense proton beams of up to 800 MeV. Based on this feasibility study, the next step will be to conduct the detailed design and construction of the prototype VPAC, as a proof-of-concept demonstration and to advance the several important applications of this accelerator.

## ACKNOWLEDGMENTS

The authors are thankful to Dr. Andrew J. Jason at the U.S. Los Alamos National Laboratory for several valuable discussions and advice on the accelerator's modeling and to Dr. Lloyd M. Young on RFQ injectors.

**Table 1. Key Parameter Values of the Variable-Phase AC Prototype Accelerator**

| PARAMETER | UNIT | VALUE |
|---|---|---|
| Beam Specie | | Proton |
| $E_{inj}$ Injected RFQ Beam Energy | MeV | 2.0 |
| $E_{ext}$ Extracted VPAC Beam Energy | MeV | 25.6 |
| $R_{inj}$ Beam Injection Radius | m | 2.5 |
| $R_{ext}$ Beam Extraction Radius | m | 5.5 |
| I Accelerated Beam Average Current | mA | 100 |
| $\Delta E_s / E_{ext}$ HWHM Beam Energy Spread | % | 0.25 |
| $N_c$ Number of Acceleration Cavities | | 4 |
| $N_s$ Number of Sectors per Turn | | 8 |
| $S_n$ Beam Path Length in Sectors | m | 3.8 – 9.65 |
| $L_m$ Length of Sector Bending Magnets | m | 0.34 – 1.0 |
| H Field Strength in Sector Magnets | T | 0.39 – 0.72 |
| G Gradient in Quadrupole Lenses | T/m | 3 – 27 |
| $\Delta E$ Energy Gain per Turn | MeV | 1.4 – 3.2 |
| $\Delta R$ Orbit Turn-to-Turn Separation | m | 0.24 – 0.34 |
| n Number of Turns | | 10 |
| h Harmonic Number | | 36 – 28 |
| f rf Frequency | MHz | 50 |
| $e_{xy}^i$ Injected Transverse Emittance | mm-mrad | $25.0\,p$ |
| $e_L^i$ Injected Longitudinal Emittance | deg-MeV | $1.0\,p$ |
| $e_{xy}^e$ Extracted Transverse Emittance | mm-mrad | $24.8\,p/6.9\,p$ |
| $e_L^e$ Extracted Longitudinal Emittance | Deg-MeV | $0.98\,p$ |
| $x_{inj}/y_{inj}$ Horizontal/Vertical Full Beam Size at Injection | mm | 20.0/10.0 |
| $z_{inj}$ Injected Bunch Full Length | mm | 116.26 |
| $x_{ext}/y_{ext}$ Horizontal/Vertical Full Beam Size at Extraction | mm | 5.0/5.0 |
| $z_{ext}$ Extracted Bunch Full Length | mm | 112.62 |
| Vacuum Chamber Horizontal/Vertical Full Size | mm | 140/70 |

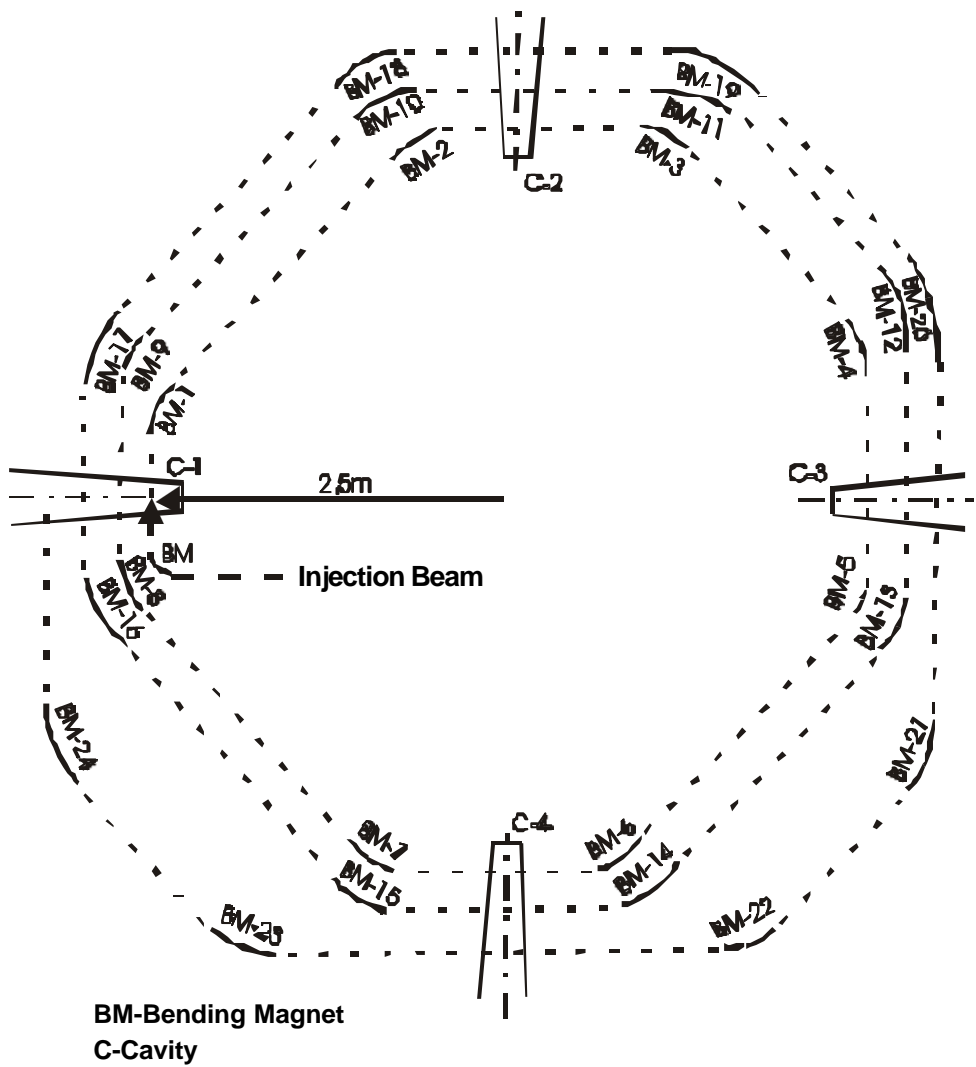

**BM-Bending Magnet**
**C-Cavity**

**Fig. 1 Trajectory of central particle in the VPAC prototype for the first three turns**

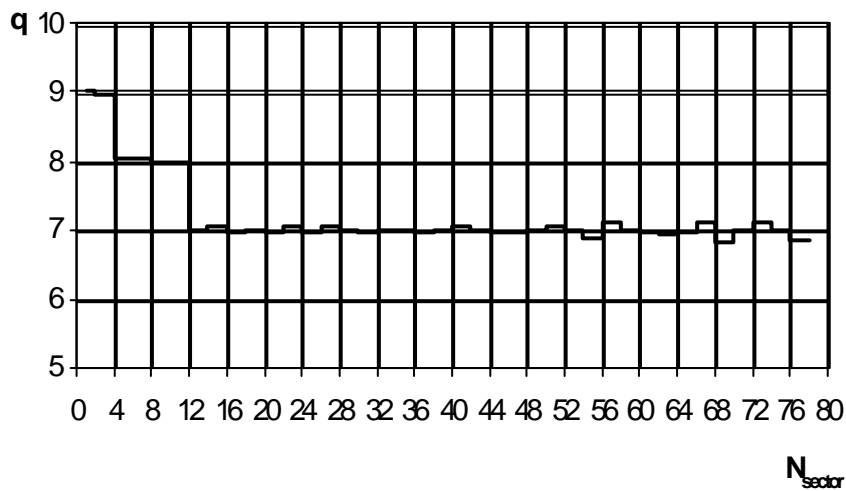

**Fig. 2 Intercavity harmonic number**

Fig. 3 TRACE 3D output of beam dynamics for
the 1-st turn of the VPAC prototype

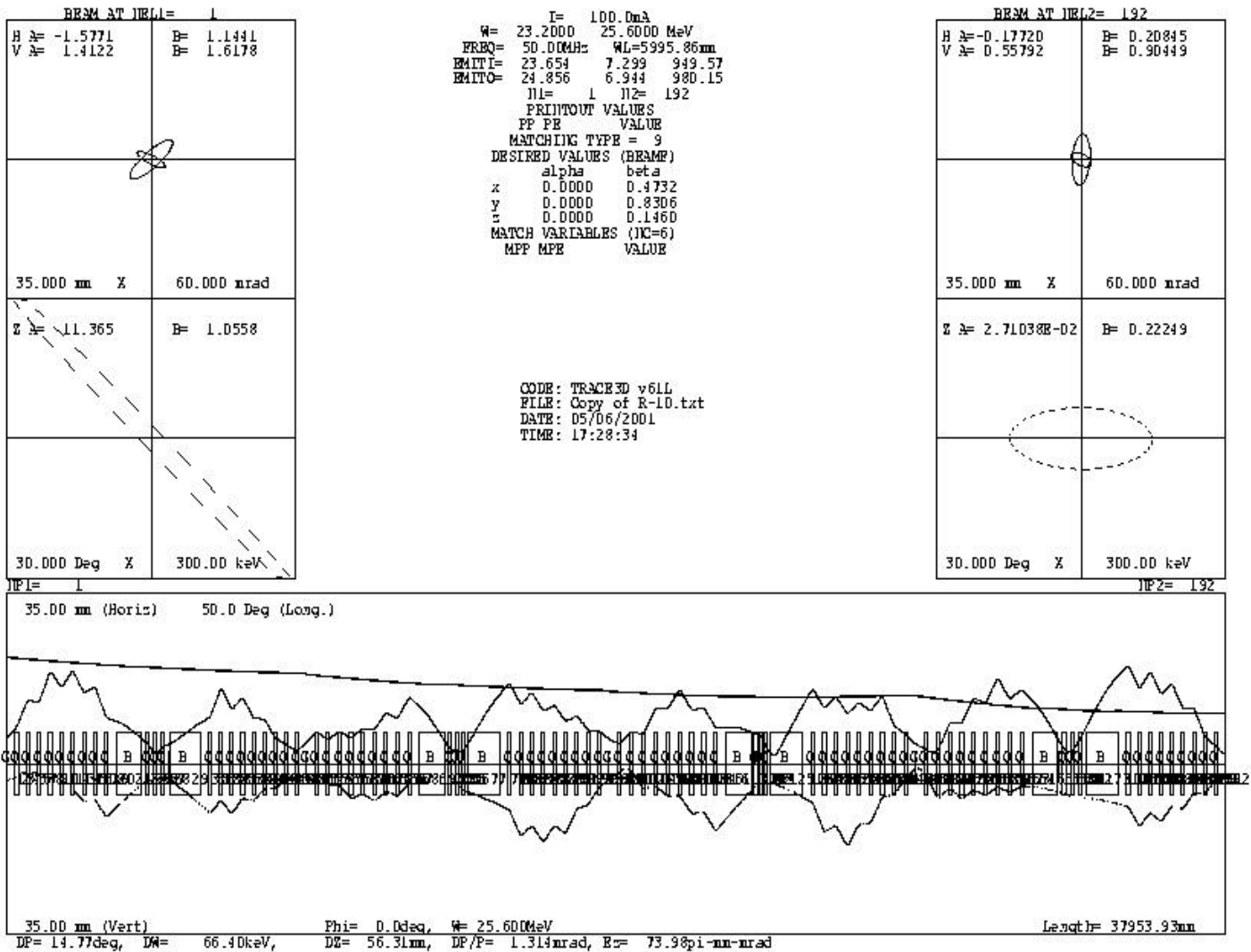

**Fig. 4 TRACE 3D output of beam dynamic for the 10-th turn of the VPAC prototype**

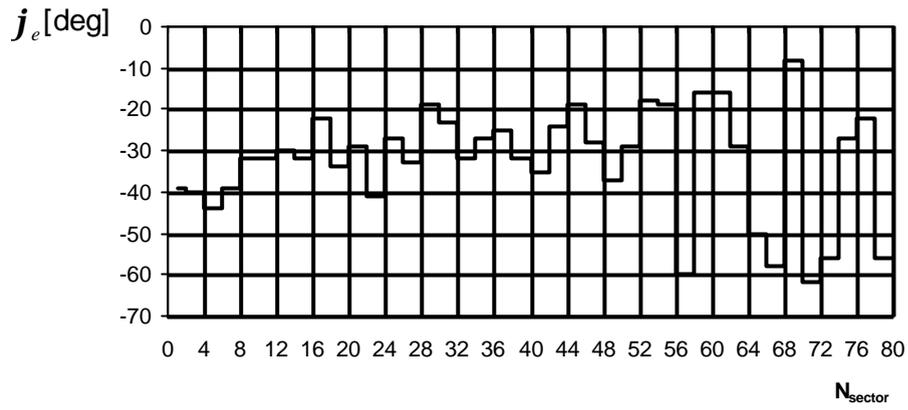
**Fig. 5 Acceleration equilibrium phase**

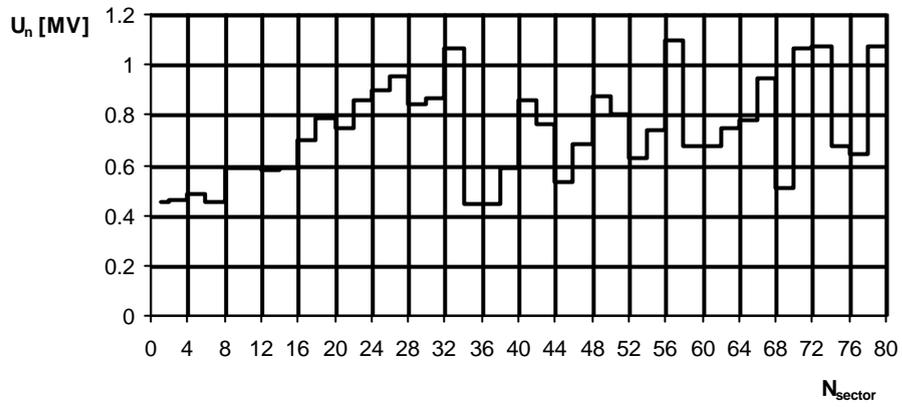
**Fig. 6 Acceleration voltage amplitude**

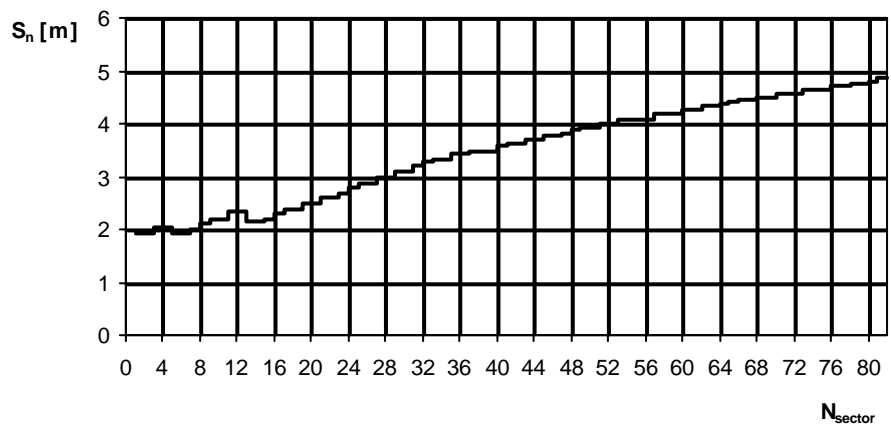
**Fig. 7 Beam trajectory length**

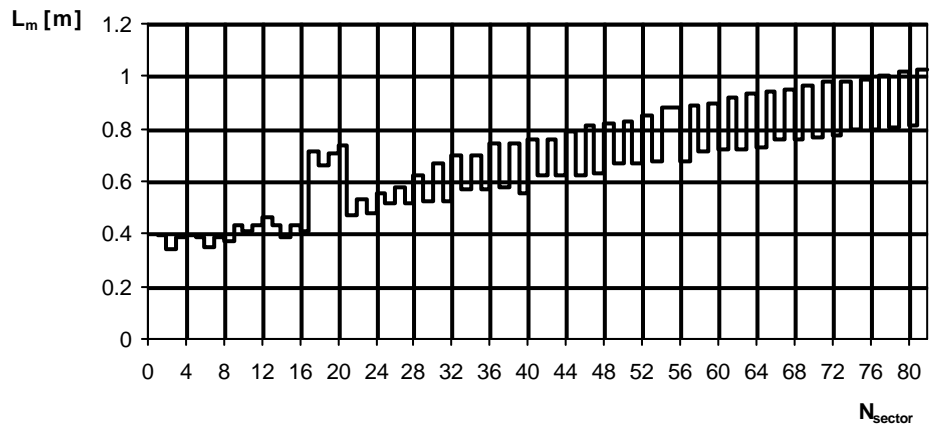

**Fig. 8 Length of sector magnets**

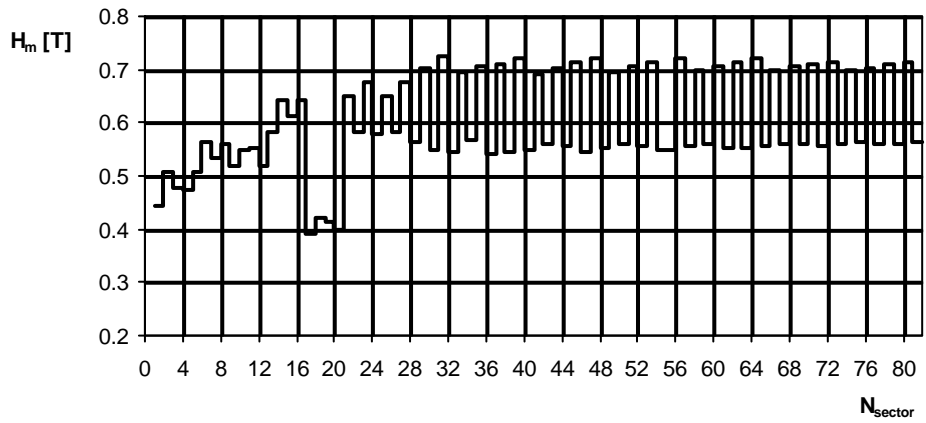

**Fig. 9 H-field in sector magnets**

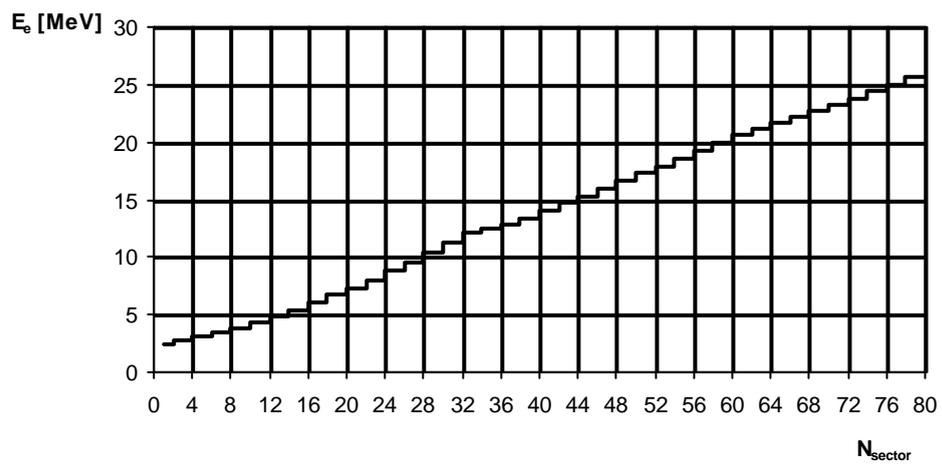

**Fig. 10 Kinetic energy**

15